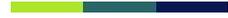

ARTICLE

# Predicting the properties of molecular materials: multiscale simulation workflows meet machine learning

Fabio Le Piane[1], Matteo Baldoni[1] and Francesco Mercuri[1]*

**Date:** 29 May 2020

Machine Learning tools are nowadays widely applied extensively to the prediction of the properties of molecular materials, using datasets extracted from high-throughput computational models. In several cases of scientific and technological relevance, the properties of molecular materials are related to the link between molecular structure and phenomena occurring across a wide set of spatial scales, from the nanoscale to the macroscale. Here, we describe an approach for predicting the properties of molecular aggregates based on multiscale simulations and machine learning.

[1]DAIMON Team, Consiglio Nazionale delle Ricerche (CNR), Istituto per lo Studio dei Materiali Nanostrutturati (ISMN), Via P. Gobetti 101, Bologna, Italy. Email: francesco.mercuri@cnr.it

# Introduction

In recent years, machine learning (ML) methods have applied with success to studies of the properties of molecular materials.[1–7] The vast majority of these studies are focused on the properties of individual molecules, targeting the correlation between molecular structure and resulting properties.[8–10] The properties of several technological materials constituted by molecular aggregates, however, depend on both molecular structure and on aggregation morphology, as for example in the case of nanoscale materials.[11,12] Computational methods for predicting the properties of molecular materials must therefore integrate the properties of individual molecules with information about aggregation morphology, which, in turn, can be related to materials fabrication and processing.[13] The definition of a modelling paradigm able to simulate and predict the properties of molecular materials as a function of molecular structure and aggregation/fabrication conditions can potentially enable high-throughput development of novel materials for technological applications.

In this work, we design and implement a computational workflow for the simulation of the properties of molecular materials integrated with a ML scheme for enhancing the computational workload. The workflow is based on a multi-scale top-down approach, in which target properties are defined from the application to the molecular scale. The workflow is implemented through top-down hierarchical data structures, which connects the properties of molecular materials at the nanoscale to the atomistic/electronic scale. Modelling data are generated by applying domain-specific simulation protocols based on atomistic molecular dynamics (MD) and density functional theory (DFT) calculations. ML approaches are therefore applied to enable the scale reduction, providing a local mapping at a lower scale of the properties of large molecular aggregates, reducing greatly the overall computational load.

The proposed approach is applied to the evaluation of intermolecular electronic couplings of aggregates of organic molecular semiconductors, a key quantity for the development of materials for advanced electronics. Preliminary studies suggest the relevance of the specific set of features considered for representing intermolecular properties, which depend on the aggregation morphology. Work is in progress to assess the interplay between the structure of individual molecules and the structure of aggregates in determining the performance of ML predictions of the properties of molecular materials. Moreover, an additional speedup of the whole workflow is obtained by optimizing the implementation of the integration between the multiscale simulation workflow and the ML engine.

## Molecular materials: from individual molecules to aggregates and interfaces

Generally speaking, the properties of molecular materials depend both on the properties of individual molecules (related to the chemical composition and structure of a molecule) and on the properties of aggregates. For example, in Fig. 1 the complex structure of the morphology obtained from molecular dynamics simulations of aggregation of a perylene diimide derivative is shown.[14,15]

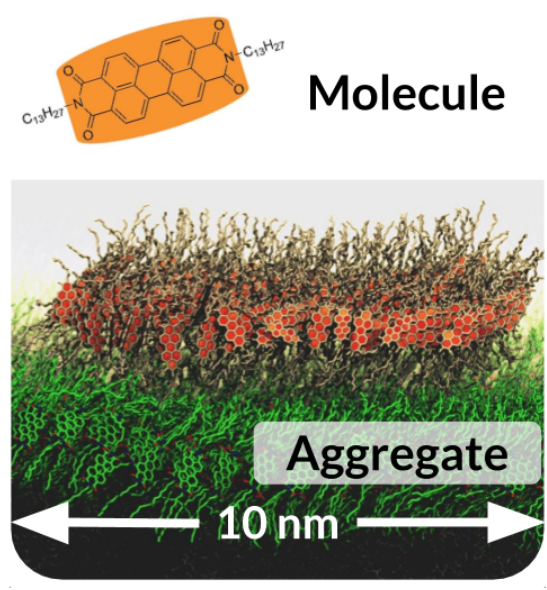

**Figure 1.** Molecular structure of a perylene diimide derivative and resulting simulated morphology at the interface with a substrate.

The aggregation morphology depends, in turn, on both the peculiar molecular structure and on processing conditions and environment.[16] In several cases of technological interest, the resulting aggregate exhibits structural features on the nanometer length scale. Indeed, nanoscale aggregation and morphology impact on several properties of molecular materials.[17] The evaluation and prediction of the properties of aggregate must therefore consider the properties of materials across a quite wide range of length scales, from the molecular scale to the nano- and micro-scale.

Multiscale simulations techniques provide tools for the modelling of the properties of materials at different scales.[18,19] In the particular cases considered in this work, multiscale simulations can be used to link the properties of individual molecules to the properties of molecular aggregates. Specifically, different computational methods target phenomena occurring at different scales, and the output of a simulation at a given scale can be used as an input to perform another set of simulations at a lower or higher scale, providing the cross-scale link.

## Charge transport in molecular semiconductors

A particularly interesting case study concerns the evaluation and prediction of the charge transport properties in molecular semiconductors. The charge transport properties of molecular materials are exploited in several cases of technological interest, for example in the development of organic light-emitting diodes (OLEDs) or organic photovoltaic (OPV) solar cells. In several cases, the propensity to efficient charge transport depends on the intrinsic electronic properties of materials, as for example occurring in functionalized carbon-based nanostructures.[20–24] In the case of molecular materials, however, the overall properties of the materials, in terms of phenomena related to charge transport, depend on i) the electronic properties of individual molecules (electronic configuration, energy levels, etc.) and ii) molecular aggregation, intermolecular interactions, morphology, deformations, interfaces and all other effects concerning the interaction of individual molecules with other molecules or materials.[13]

A very simple, though effective, model of molecular semiconductors describes the charge transport process in terms of percolation of charge by **hopping** from a molecule towards a neighbouring molecule, as shown in Fig. 2.

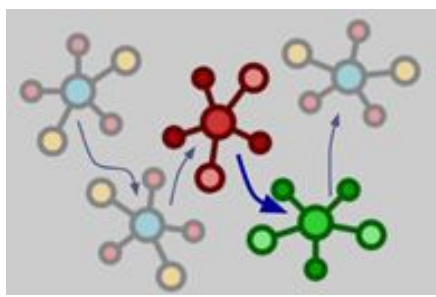

**Figure 2.** Charge hopping mechanism in molecular semiconductors.

The current flowing through materials can essentially be interpreted as a sequence of single events involving two neighboring molecules. A key quantity for the evaluation of charge transport in organic semiconductors is therefore the electronic coupling between two neighboring molecules. This quantity can usually be determined by DFT simulations, involving (in the most simple case and neglecting collective effects) pairs of molecules. The transport properties of molecular aggregates can subsequently be obtained by a sort of weighted statistical integration, for example by applying kinetic Monte Carlo (kMC) simulations.[25,26] However, two relevant issues must be considered:

1. The evaluation of the intermolecular couplings by DFT simulations is quite demanding, from the computational side, and can require up to a few CPU hours, for a single molecular pair, on standard computational infrastructures.

2. As we discussed before, the properties of molecular materials, including charge transport, depend strongly on the aggregation morphology and on resulting interactions on a scale of several tens or hundreds of nanometers. As the typical size of individual molecules is on the order of a few nanometers, the evaluation of intermolecular coupling in nanoscale aggregates results in several thousands of pairs, each of which needs an independent DFT calculation.

The use of statistical methods, such as kMC, for the evaluation of charge transport properties requires a balance between accuracy and computational load, which can exceed several thousands of CPU hours (see Fig. 3).

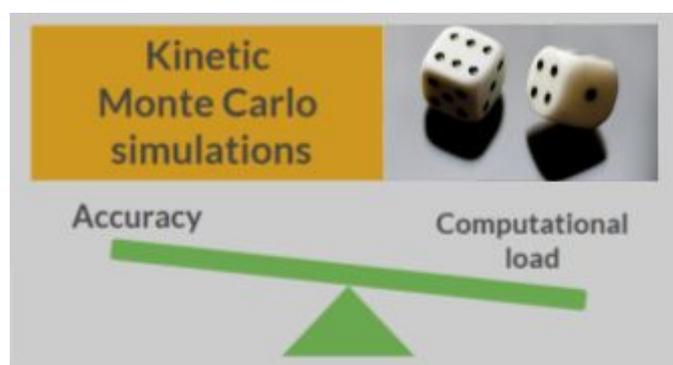

**Figure 3.** The application of kMC methods relies on statistical accuracy, which results in high computational loads.

We need therefore a set of tools which can assist the evaluation of the charge transport properties of molecular materials with good accuracy and, possibly, saving CPU time.

## A multiscale top-down approach: from simulations to data workflows

Our approach relies on a top-down view of the properties of molecular materials for applications. For example, we can consider the properties of active materials used in organic electronic devices as derived from interlinked materials properties on progressively lower length scales, from the device to the molecular scale, as shown in Fig. 4.

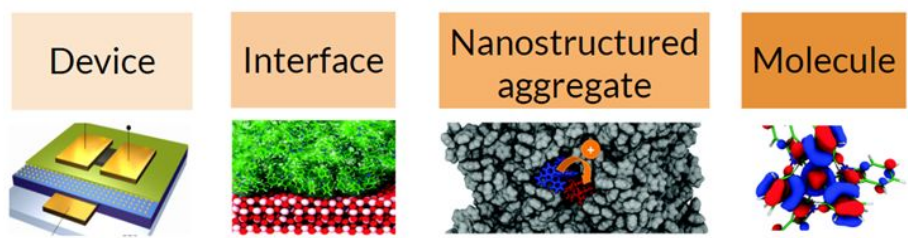

**Figure 4.** Top-down description of the properties of active materials used in organic electronic devices. Partially adapted from Ref. [27] [28] [13,28] © 2010 American Chemical Society and with permission from the RSC.

In this case, we can first consider the aggregation morphology of molecular materials at the nanoscale. For example, we can simulate the aggregation of molecules, in different conditions, by atomistic (or coarse-grained) MD.[29] This step will also link the nanoscale morphology of molecular materials to processing or fabrication conditions, a fundamental part in the engineering of organic electronic devices.[30] Then, we can proceed to a reduction of the scale, extracting pairs of neighboring molecules from the MD configurations and computing electronic couplings for each pair (see Fig. 5). As explained before, however, this step may require the evaluation of electronic coupling for a large number of molecular pairs, in the order of thousands or more.

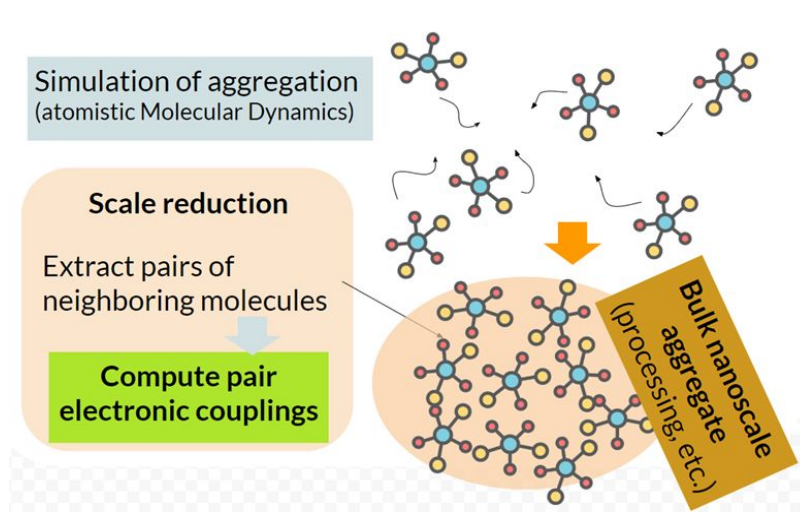

**Figure 5.** Simulation of the aggregation morphology of molecular materials by MD, from which individual pairs are extracted for subsequent DFT calculations.

It is worth noting that the top-down approach discussed above relies, technically, on the knowledge of the molecular structure only. Indeed, the aggregation morphology of molecular materials, at least for pure bulk materials, depends on the molecular structure and aggregation conditions only. The whole process of pair selection can therefore be represented as in Fig. 6.

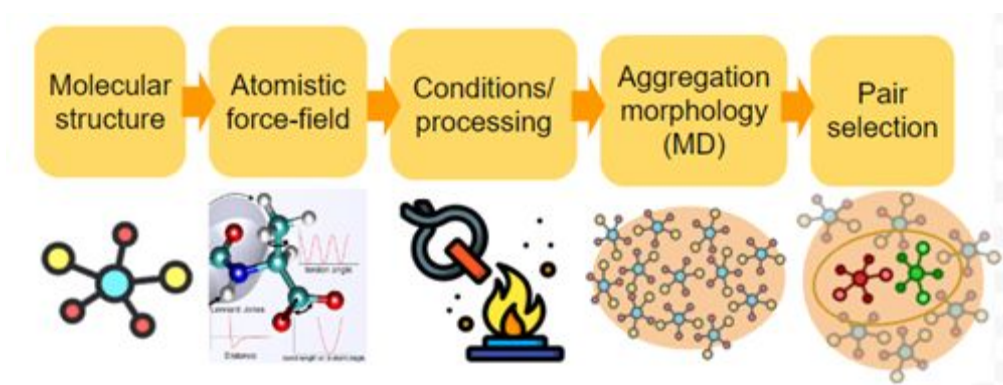

**Figure 6.** Multiscale workflow for the simulation of charge transport properties in molecular aggregates.

We start from the knowledge of the structure of an individual molecule. On the basis of this knowledge, we build a suitable atomistic potential, usually in terms of a force field, including intramolecular and intermolecular terms. We select the conditions leading to aggregation and build a computational model that is able to reproduce the aggregation morphology by MD simulations. The individual molecular pairs are extracted from the simulated aggregate, and DFT calculations are carried out for each selected pair. This set of steps also defines a flow of data which links molecular structure to charge transport properties.

## Learning charge transport properties from simulated data

The last step of the workflow shown in Fig. 6 suggests that a statistically accurate method for the prediction of the properties of molecular pairs, based on a representative set of selected pairs, can greatly enhance the computational performance of the approach proposed (see Fig. 7). The gain in computational efficiency is related to the very large number of molecular pairs in a nanoscale aggregate, for each of which electronic coupling must be evaluated. Nevertheless, in the theoretical framework considered, the intermolecular electronic coupling depends on the atomic positions of all atoms in the pair only, which can be simply derived from simulations at a higher scale.

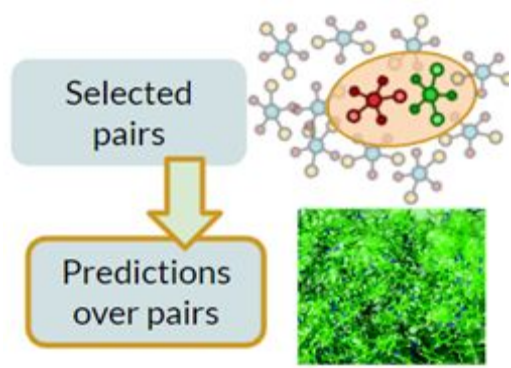

**Figure 7.** Selection of molecular pairs as datasets for automatic prediction and extrapolation over a larger number of pairs.

In other words, we can use the pair/property relationship, learned from data simulated on a selected set of pairs, to extend predictions to a very large number of pairs, thus overcoming the computational bottleneck related to the application of DFT methods.[28]

We can therefore figure out a link between the multiscale framework, applied for the evaluation of the properties of molecular materials, and a possible application of a ML framework, to enhance the overall computational throughput. In particular, we observe that a molecular aggregate constituted by structurally identical molecules can be described as a set of molecules that are translated, rotated and distorted with respect to a reference molecule (see Fig. 8). In terms of ML models, a feature describing a molecular pair, which is our target object, must describe the structure of a given pair of molecules within an aggregate. As in our computational model we are neglecting collective effects, the link between the structure of a molecular pair and the target property (intermolecular charge coupling) can be defined in terms of the relative configuration of the two molecules, in which one of the two molecules is roto-translated with respect to the other, and both molecules are possibly distorted.

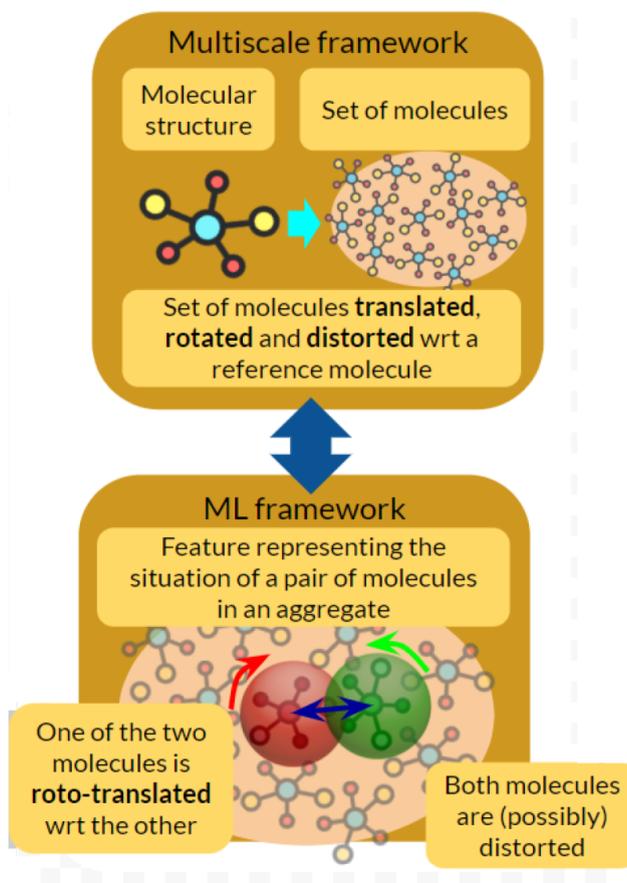

**Figure 8.** Connection between the multiscale framework, for the evaluation of the morphology in molecular aggregates, and the ML framework, defining features that describe molecular pairs in aggregates.

The features considered in the ML models must therefore be representative of the situation of each of the two molecules and of the molecular pair in an aggregate. The features can for example include quantities related to the translation, rotation and distortion of a molecule with respect to the other, and can be defined in terms of representations derived from the quaternion algebra or from the (full or simplified) intermolecular Coulomb matrix, just to mention a few. The combination of translation vectors and quaternion rotations, in particular, provide an efficient way to encode the relative position in space of two objects.[31,32] Moreover, in the definition of features related to the representation of molecular pairs, molecular symmetry can be used to derive simplified descriptors or to augment the original dataset of simulated quantities (see Fig. 9).

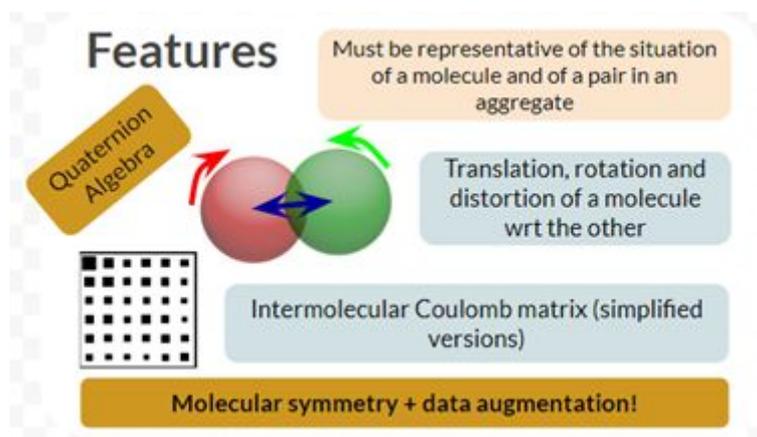

**Figure 9.** Extraction of features to represent molecular pairs in aggregates.

The predictive framework can be implemented by applying ML models to the simulated set of data, using the features described above. In practical applications, we tested a wide range of predictive models, including random forest, kernel ridge regression and deep neural networks. The assessment of the efficiency of convolutional neural networks is under investigation. We implemented the whole workflow making use of tools for the development of ML frameworks for molecular sciences based on Python, such as scikit-learn,[33] MD analysis,[34] RDKit,[35] and several others. Work is in progress for the development of a general-purpose framework for the application of ML models to molecular sciences based on Rust[36] and Julia.[37]

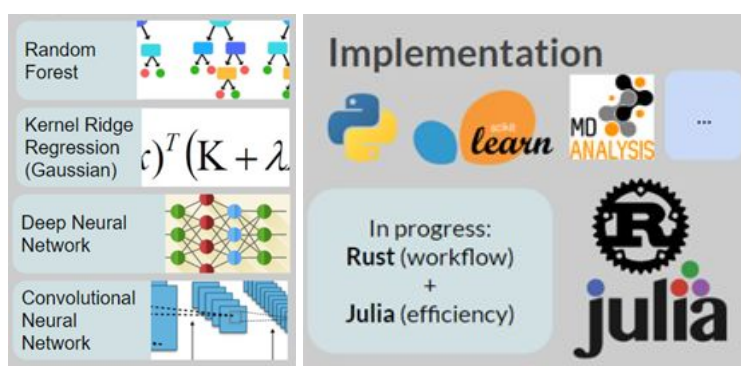

**Figure 10.** ML models and implementation tools.

Preliminary implementations have already demonstrated a significant boost in performance related to the use of Julia. The details of the Rust/Julia implementation will be discussed in further work.

## Case study and numerical results

We applied the approach proposed to a case study of scientific and technological relevance, that is, the evaluation of charge transport properties of a molecular semiconductor material used in OLEDs.[38,39] Essentially, we applied the workflow depicted in Fig. 6 to the case of a prototypical semiconductor molecule, here code-named DPBIC.[28] MD calculations were first performed to simulate the bulk morphology of amorphous molecular aggregates, thus obtaining a realistic model of the configuration occurring in typical fabrication conditions. As explained above, the accuracy of MD simulations relies on a suitable definition of the intermolecular and intramolecular interaction potential. From the MD trajectory, random snapshots of the equilibrated configuration were extracted, and a set of nearest-neighbour molecular pairs was selected. For each pair, the intermolecular electronic coupling was computed by DFT. On the basis of the dataset obtained from multiscale (MD + DFT) simulations, corresponding to around 2000 pairs, we trained ML models using different sets of features. Some preliminary results are shown in Fig. 11.

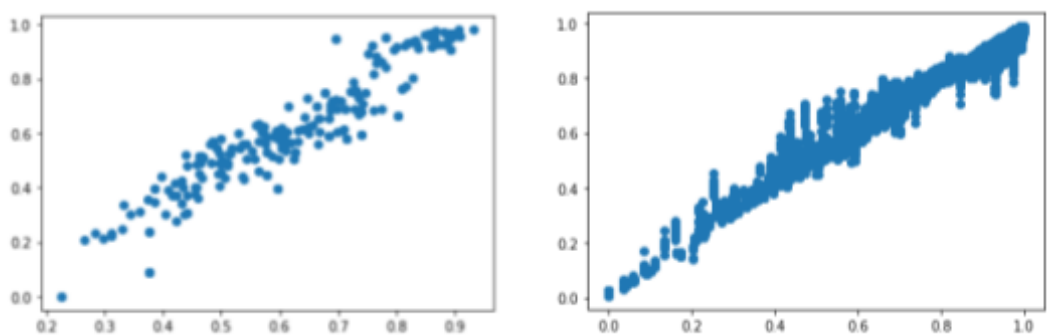

**Figure 11.** Normalized true (x) and predicted (y) intermolecular electronic coupling (log scale) obtained using gradient boosting and the intermolecular Coulomb matrix as a feature (left) and full cartesian coordinates and by applying a kernel ridge model (right).

Several details of the ML models must still be optimized, including also the selection of a feature vector able to represent efficiently the configuration of molecular pairs, which can be considered as a set of two molecules in space, with same chemical composition and structure and local distortions, affecting the relative position of atoms. However, the preliminary results obtained show that quantitatively predictive models of the intermolecular electronic coupling can indeed be obtained from a relatively small dataset, with good accuracy over a wide range of values and intermolecular configurations. In particular, our results demonstrate that, in the case of the molecular system considered, molecular deformations take a significant role in determining the intermolecular electronic coupling. In other words, the intermolecular configuration originating the molecule-molecule coupling can be reduced to a geometrical relationship (translation and relative rotation) between two rigid objects only to a first approximation. Most importantly, the predictive model can be applied to evaluate the electronic couplings in large aggregates (in the order of hundreds of nanometers or more), thus reducing the overall computational load by several orders of magnitude (see Fig. 12).

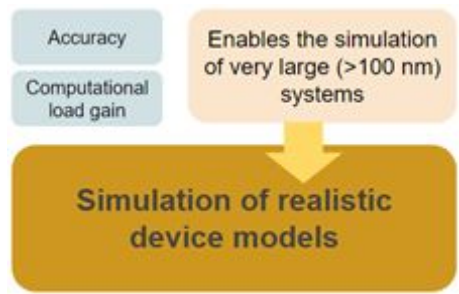

**Figure 12.** The gain in the computational load at the same level of accuracy, enabled by ML, can lead to the simulation of very large systems, providing realistic models of complex devices.

## Conclusions

The integration between multiscale simulation workflows and ML methods can greatly enhance the overall throughput of predictive models of complex devices and systems based on molecular materials. One of the key points for obtaining efficient and predictive models consists in the analysis of the phenomena, at different scales, related to the target properties, and to the consequent definition of the features to be implemented in ML models. This step also allows, in principle, a seamless integration between the computational modelling activities and the ML platform, leading to a consistent and efficient data-driven workflow. A case study, targeted to the development of a predictive model for the evaluation of charge transport properties in molecular materials, indicates that the approach proposed has the potential to improve the overall computational throughput by several orders of magnitude with respect to traditional full-scale approaches. The improvement expected by coupling multiscale simulations with ML can therefore enable the simulation of very large systems, leading to realistic models of complex devices of technological relevance.